\documentstyle[psfig]{mn}

\def\degr{\hbox{$^\circ$}}
\def\>{$>$}
\def\<{$<$}
\def\simlt{\lower.5ex\hbox{$\; \buildrel < \over \sim \;$}}
\def\simgt{\lower.5ex\hbox{$\; \buildrel > \over \sim \;$}}

\newif\ifAMStwofonts

\def\simlt{\lower.5ex\hbox{$\; \buildrel < \over \sim \;$}}
\def\simgt{\lower.5ex\hbox{$\; \buildrel > \over \sim \;$}}

\ifoldfss
  \ifCUPmtlplainloaded \else
    \NewTextAlphabet{textbfit} {cmbxti10} {}
    \NewTextAlphabet{textbfss} {cmssbx10} {}
    \NewMathAlphabet{mathbfit} {cmbxti10} {} 
    \NewMathAlphabet{mathbfss} {cmssbx10} {} 
  \fi
  \ifAMStwofonts
    \ifCUPmtlplainloaded \else
      \NewSymbolFont{upmath} {eurm10}
      \NewSymbolFont{AMSa} {msam10}
      \NewMathSymbol{\upi}     {0}{upmath}{19}
      \NewMathSymbol{\umu}     {0}{upmath}{16}
      \NewMathSymbol{\upartial}{0}{upmath}{40}
      \NewMathSymbol{\leqslant}{3}{AMSa}{36}
      \NewMathSymbol{\geqslant}{3}{AMSa}{3E}

    \fi
  \fi
\fi 

\ifnfssone
  \newmathalphabet{\mathit}
  \addtoversion{normal}{\mathit}{cmr}{m}{it}
  \addtoversion{bold}{\mathit}{cmr}{bx}{it}
  \newmathalphabet{\mathbfit} 
  \addtoversion{normal}{\mathbfit}{cmr}{bx}{it}
  \addtoversion{bold}{\mathbfit}{cmr}{bx}{it}
  \newmathalphabet{\mathbfss} 
  \addtoversion{normal}{\mathbfss}{cmss}{bx}{n}
  \addtoversion{bold}{\mathbfss}{cmss}{bx}{n}
  \ifAMStwofonts
    \ifCUPmtlplainloaded \else
      \UseAMStwoboldmath
      \makeatletter
      \new@mathgroup\upmath@group
      \define@mathgroup\mv@normal\upmath@group{eur}{m}{n}
      \define@mathgroup\mv@bold\upmath@group{eur}{b}{n}
      \edef\UPM{\hexnumber\upmath@group}
      \new@mathgroup\amsa@group
      \define@mathgroup\mv@normal\amsa@group{msa}{m}{n}
      \define@mathgroup\mv@bold\amsa@group{msa}{m}{n}
      \edef\AMSa{\hexnumber\amsa@group}
      \makeatother
      \mathchardef\upi="0\UPM19
      \mathchardef\umu="0\UPM16
      \mathchardef\upartial="0\UPM40
      \mathchardef\leqslant="3\AMSa36
      \mathchardef\geqslant="3\AMSa3E
    \fi
  \fi
\fi 

\ifnfsstwo
  \DeclareMathAlphabet{\mathbfit}{OT1}{cmr}{bx}{it}
  \SetMathAlphabet\mathbfit{bold}{OT1}{cmr}{bx}{it}
  \DeclareMathAlphabet{\mathbfss}{OT1}{cmss}{bx}{n}
  \SetMathAlphabet\mathbfss{bold}{OT1}{cmss}{bx}{n}
  \ifAMStwofonts
    \ifCUPmtlplainloaded \else
      \DeclareSymbolFont{UPM}{U}{eur}{m}{n}
      \SetSymbolFont{UPM}{bold}{U}{eur}{b}{n}
      \DeclareSymbolFont{AMSa}{U}{msa}{m}{n}
      \DeclareMathSymbol{\upi}{0}{UPM}{"19}
      \DeclareMathSymbol{\umu}{0}{UPM}{"16}
      \DeclareMathSymbol{\upartial}{0}{UPM}{"40}
      \DeclareMathSymbol{\leqslant}{3}{AMSa}{"36}
      \DeclareMathSymbol{\geqslant}{3}{AMSa}{"3E}
    \fi
  \fi
\fi 

\ifCUPmtlplainloaded \else
  \ifAMStwofonts \else 
    \def\upi{\pi}
    \def\umu{\mu}
    \def\upartial{\partial}
  \fi
\fi

\title[Eclipse maps of spiral shocks in IP~Pegasi]
		{Eclipse maps of spiral shocks in the accretion disc of 
		IP~Pegasi in outburst}
\author[R. Baptista et~al.]
       {Raymundo Baptista$^1$, E. Harlaftis$^2$ and D. Steeghs$^{3,4}$ \\
       $^1$ Departamento de F\'\i sica, Universidade Federal de Santa Catarina,
       Campus Trindade, 88040-900, Florian\'opolis - SC, Brazil, \\
       ~ email: bap@fsc.ufsc.br \\
       $^2$ Astronomical Institute, Observatory of Athens, Lofos Koufou,
       P.\ Penteli, Athens, 152 36, Greece, email: ehh@astro.noa.gr \\
       $^3$ School of Physics \& Astronomy, University of St.\,Andrews,
       North Haugh, St.\,Andrews, Fife, KY16 9SS, Scotland \\
       $^4$ Physics \& Astronomy, University of Southampton, Highfield,
       Southampton, SO17 1BJ, UK, email: ds@astro.soton.ac.uk }
\date{Accepted for publication at Monthly Notices of the Royal Astronomical
		Society}

\pagerange{1--7}
\pubyear{2000}

\begin{document}

\label{firstpage}

\maketitle

\begin{abstract}

Eclipse lightcurves of the dwarf nova IP Peg during the November
1996 outburst are analysed with eclipse mapping techniques to 
constrain the location and investigate the spatial structure of the
spiral shocks observed in the Doppler tomograms (Harlaftis et~al. 1999).
Eclipse maps in the blue continuum and in the C\,III+N\,III
$\lambda 4650$ emission line show two asymmetric arcs of $\sim 90$
degrees in azimuth and extending from intermediate to the outer disc
regions ($R\simeq 0.2 - 0.6\; R_{L1}$, where $R_{L1}$ is the distance
from disc centre to the inner Lagrangian point) which are interpreted 
as being the spiral shocks seen in the Doppler tomograms.
The He\,II $\lambda 4686$ eclipse map also shows two asymmetric arcs
diluted by a central brightness source.
The central source probably corresponds to the low-velocity
component seen in the Doppler tomogram and is understood in terms
of gas outflow in a wind emanating from the inner parts of the disc.
We estimate that the spirals contribute about 16 and 30 per cent of
the total line flux, respectively, for the He\,II and C\,III+N\,III lines.
Comparison between the Doppler and eclipse maps reveal that the 
Keplerian velocities derived from the radial position of the shocks are systematically larger than those inferred from the Doppler tomography
indicating that the gas in the spiral shocks has sub-Keplerian velocities.
We undertake simulations with the aim to investigate the effect of
artifacts on the image reconstruction of the spiral structures.

\end{abstract}


\section{Introduction}

Accretion discs are widespread in astrophysical environments, from
sheltering the birth of stars to providing the energetics for the most
violent phenomena such as relativistic jets.
Despite its general importance and although considerable
effort both in observation and theory has been invested over the past
decade, the structure and underlying physics of accretion discs remains
poorly understood. One of the major unsolved problems concerns the
nature of the anomalous viscosity mechanism responsible for the
inward spiraling of the disc material (Frank, King \& Raine 1992).

Best prospects for progress in understanding accretion discs physics
are possibly found in mass-exchanging binaries such as Cataclysmic
Variables (CVs). In these close binaries mass is fed to a non-magnetic
($B\simlt 10^{5}$ G) white dwarf via an accretion disc by a Roche lobe
filling companion star (the secondary). The sub-class of 
{\em dwarf novae} comprises low-mass transfer CVs which show recurrent
outbursts of 2--5 magnitudes on timescales of months either due to an
instability in the mass transfer from the secondary or due to a thermal
instability in the accretion disc which switches the disc from a low 
to a high-viscosity regime (Warner 1995 and references therein).

Spiral shocks have been advocated by various researchers as a possible
mechanism for transport of angular momentum in accretion discs (Savonije,
Papaloizou \& Lin 1994) and may be the key, together with magnetic
viscosity (Hawley, Balbus \& Winters, 1999), in understanding the
viscosity mechanism.
The recent discovery of spiral shocks in the accretion disc of the
dwarf novae IP~Pegasi in outburst -- from Doppler tomography of emission
lines (Steeghs, Harlaftis \& Horne 1997, 1998; Harlaftis et~al. 1999) 
-- confirmed the results of hydrodynamical simulations (Armitage \& Murray
1998, Stehle 1999).
The spiral shocks are produced in the outer regions of the disc by the
tides raised by the secondary star. During the outburst the disc expands
and its outer parts feel more effectively the gravitational attraction
of the secondary star leading to the formation of spiral arms.

Here we report on the eclipse mapping analysis of the data obtained by
Harlaftis et al. (1999; see there for observations and data reduction).
Our goal is to confirm the existence, constrain the location and to
investigate the spatial structure of the spiral shocks observed in
the Doppler tomograms. Section\,\ref{dados} presents the data and
gives details of the analysis procedures. In section\,\ref{sim} we
present a set of simulations with the eclipse mapping method aimed to
clarify the interpretation of the results of section\,\ref{results} in
terms of real spiral shocks. A summary of our findings is given in 
section\,\ref{fim}.

\section{Data Analysis} \label{dados}

\subsection{Lightcurves}

A time-series of high-resolution, optical spectrophotometry 
($\Delta\lambda= 4354-4747$ \AA, velocity dispersion of $27\; km\,s^{-1}$
per pixel) covering one eclipse of IP Peg was obtained during the third
day of the November 1996 outburst. The reader is referred to Harlaftis 
et~al. (1999) for a detailed description of the dataset and of the
reduction procedures.
Lightcurves were extracted for the blue continuum ($4365-4440$ \AA)
and for the C\,III+N\,III $\lambda 4650$ (Bowen blend) and He\,II $\lambda
4686$ lines and phase folded according to the sinusoidal ephemeris of 
Wolf et~al. (1993),
\begin{equation}
T_{\rm mid}({\rm HJD}) = 2\,445\,615.4156 + 0.158\,206\,16 \; E + (O-C) 
\end{equation}
where 
$$(O-C) = 1.0903 \times 10^{-3} \; \sin \left[ 2\,\pi\;
\frac{E-10258}{10850.88} \right] \;\; . $$
The line lightcurves were continuum subtracted and, therefore,
correspond to {\em net} line emission.
The three lightcurves are shown in Fig.\,\ref{fig1} as gray open squares. 

The C\,III+N\,III lightcurve shows a peculiar double-stepped eclipse
shape revealing the presence of two asymmetric brightness sources
displaced from disc centre. Although less pronounced,
the same morphology can also be seen in the continuum lightcurve.
The shape of the He\,II eclipse is more symmetrical than that of
the other passbands but mid-eclipse occurs earlier with respect to the
continuum eclipse, indicating that the line surface distribution is 
also asymmetric.

The continuum and He\,II lightcurves show a conspicuous orbital
modulation with maximum at phase $\phi \simeq -0.15$ cycle and minimum
at $\phi \simeq +0.15$ cycle. This is not seen in the C\,III+N\,III 
lightcurve although an increase in flux is clearly visible after phase
$\phi= +0.22$ cycle. For any reasonable mass ratio, $q<1$ [$q$=0.5 
for IP~Peg, Wood \& Crawford (1986)], the shadow
of the secondary star covers regions outside the primary lobe for
orbital phases $|\phi|> 0.2$ cycle. Therefore, it is hard to explain
the observed modulation in terms of occultation by the secondary star
unless the eclipsed source lies outside the primary lobe. 
Thereafter, we assign the orbital modulation to gas obscuration by the
spiral arm seen at maximum between phases 0.0-0.25 cycle (see section\,\ref{results}).

Out-of-eclipse brightness changes are not accounted for by the eclipse
mapping method, which assumes that all variations in the eclipse lightcurve
are due to the changing occultation of the emitting region by the
secondary star (but see Bobinger et~al. 1996 for an example of how to
include orbital modulations in the eclipse mapping scheme). Orbital
variations were therefore removed from the lightcurves by fitting a
spline function to the phases outside eclipse, dividing the lightcurve
by the fitted spline, and scaling the result to the spline function
value at phase zero. This procedure removes orbital modulations with
only minor effects on the eclipse shape itself.
The corrected lightcurves are shown in Fig.\,\ref{fig1} as filled
circles. For the purpose of eclipse mapping analysis the lightcurves
were limited to the phase range ($-0.18,+0.28$) since data outside
of eclipse is basically used to set the out-of-eclipse level and
flickering in this phase range serves to add unnecessary noise to the fit. 
%

\subsection{Eclipse maps}

The eclipse mapping method is an inversion technique that uses 
the information contained in the shape of the eclipse to recover
the surface brightness distribution of the eclipsed accretion disc.
The reader is referred to Horne (1985), Rutten et al. (1992) and 
Baptista \& Steiner (1993) for the details of the method.

As our eclipse map we adopted a grid of $51 \times 51$ pixels
centered on the primary star with side 2~R$_{\rm L1}$, where
R$_{\rm L1}$ is the distance from the disk center to the inner
Lagrangian point.
The eclipse geometry is controled by a pair of $q$ and $i$ values.
The mass ratio $q$ defines the shape and the relative size of the
Roche lobes. The inclination $i$ determines the shape and extent of
the shadow of the secondary star as projected onto the orbital plane.
We obtained reconstructions for two sets of parameter, ($q=0.5 \; , \;
i=81\degr$) (Wood \& Crawford 1986) and ($q=0.58 \; , \; i=79.5\degr$)
(Marsh 1988), which correspond to an eclipse width of the disc
centre of $\Delta\phi= 0.086$ (Wood \& Crawford 1986;
Marsh \& Horne 1990). These combination of parameters ensure that
the white dwarf is at the center of the map. There is no perceptible
difference in eclipse maps obtained with either geometry. Hence, for
the remainder of the paper we will refer to and show the results for 
($q=0.5 \; , \; i=81\degr$).

The lightcurves were analyzed with eclipse mapping techniques to solve
for a map of the disc brightness distribution and for the flux of an
additional uneclipsed component in each passband. The uneclipsed component
accounts for all light that is not contained in the eclipse map in the
orbital plane (i.e., light from the secondary star and/or a vertically
extended disc wind). The reader is referred to Rutten et~al. (1992) and
Baptista, Steiner \& Horne (1996) for a detailed description of and 
tests with the uneclipsed component. For the reconstructions we adopted
the default of limited azimuthal smearing of Rutten et~al. (1992), which
is better suited for recovering asymmetric structures than the original
default of full azimuthal smearing (see Baptista et~al. 1996).

Lightcurves, fitted models and grayscale plots of the resulting eclipse
maps are shown in Fig.\,\ref{fig2} and will be discussed in detail in 
section\,\ref{results}.
%

\section{Eclipse mapping simulations} \label{sim}

We performed various simulations with asymmetric sources in order
(i) to investigate how the presence of spiral structures in the
accretion disc affects the shape of the eclipse lightcurve, and
(ii) to evaluate the ability of the eclipse mapping method to 
reconstruct these structures in the eclipse maps. 

For the simulations we adopted the geometry of IP Peg ($q=0.5$ and $i=81
\degr$) and constructed lightcurves with the same signal to noise ratio
and orbital phases of the real data of section\,\ref{dados}. 
Figure\,\ref{fig3} shows the results of the simulations.
%

Asymmetric compact sources (model \#1) result in eclipse lightcurves with
rapid brightness changes at ingress/egress phases. The azimuthal smearing
effect characteristic of the eclipse mapping method is responsible for the
distortion which makes the compact sources appear `blurred' in azimuth.
Nevertheless, their radial and azimuthal locations are satisfactory 
recovered.

Brightness distributions with spiral structures (models \#2 to \#4) result
in eclipse shapes with characteristic `bulges', whose extension and location
in phase reflect the orientation and radial extent of the spiral arms.
Due to the azimuthal smearing effect these structures are reproduced in
the form of asymmetric arcs, whose maximum brightness and radial position
yield information about the orientation, position and radial extent of
the original spiral arms. The adition of a symmetric brightness
source (i.e., centred in the eclipse map, model \#5) can dilute the
presence of spiral arms. In this case the eclipse shape is smoother
and more symmetric in comparison with those of models \#2-4 and the
asymmetric arcs are less clearly visible in the eclipse map, mixing
with the brightness distribution of the central source.

These simulations show that the eclipse mapping method is able to
reproduce asymmetric light sources such as spiral arms (provided the
asymmetric sources are properly eclipsed) and that the asymmetric
structures seen in the eclipse maps of Fig.\,\ref{fig2} are not caused
by artifacts of the method. Models \#3 to \#5 are the relevant ones for
the purpose of comparing the results of the simulations with those
from the IP~Peg data (Fig.\,\ref{fig2}). The morphology of the 
continuum and C\,III+N\,III lightcurves is similar to that of models
\#3 and \#4, while the He\,II lightcurve resembles that of model \#5.

\section{Results} \label{results}

Data and model lightcurves are shown in the left panels of Fig.\,\ref{fig2}.
Horizontal dashed lines indicate the uneclipsed component in each case.
The uneclipsed component corresponds to about 12, 1 and 1 per cent of the
total flux, respectively for the continuum, C\,III+N\,III and He\,II curves.
While a non-negligible fraction of the light in the continuum 
probably arises from an emitting region outside of the orbital plane 
(possibly a disk wind), the net He\,II and C\,III+N\,III emission mostly
arises from (or close to) the orbital plane.

The middle panels of Fig.\,\ref{fig2} show eclipse maps in a logarithmic
grayscale. The maps in the right panels show the asymmetric part of the
maps in the middle panels and are obtained by calculating and subtracting
azimuthally-averaged intensities at each radius.

The continuum and C\,III+N\,III lightcurves display bulges similar to
that of the lightcurves of models \#3 and \#4 (Section 3) and result in
eclipse maps with two clearly visible asymmetric arcs, which are
interpreted as being the spiral shocks seen in the Doppler tomograms
of Harlaftis et~al. (1999). In comparison with the models of
Fig.\,\ref{fig3}, the orientation of the arcs suggests that the spirals
are aligned in a direction perpendicular to the major axis of the binary
(models \#3 and \#4). The arcs show an azimuthal extent of $\sim 90\degr$
and extend from the intermediate to the outer disc regions ($R\simeq
0.2-0.6\;R_{\rm L1}$). Therefore, the outer radius of the spirals is
of the same order of the disc outburst radius ($R_d \simeq 0.34\;a \simeq
0.6\;R_{L1}$) inferred by Wood et~al. (1989). The eclipse maps show no
evidence of the bright spot at disc rim and no enhanced emission along
the gas stream trajectory.

The azimuthal location of the arcs is consistent with the results
from hydrodynamical simulations and from the Doppler tomography.
The arc in the upper left quadrant of the eclipse map (hereafter
arc 1) corresponds to the spiral arm whose maximum occurs at phases
0.5--0.75 while the arc in the lower right quadrant (arc 2) corresponds
to the spiral arm seen at maximum intensity at phases 0.0--0.25
(orbital phases increases clockwise in the eclipse maps of 
Fig.\,\ref{fig2} and phase zero coincides with the inner lagrangian
point L1). The arcs are not symmetrical with respect to the centre of
the disc. In C\,III+N\,III, arc 2 is further away from disc centre
than arc 1 -- in agreement with the Doppler tomography, which indicates
smaller velocities for the spiral arm 2 ($\simeq 550\; km\, s^{-1}$)
than for the spiral arm 1 ($\simeq 700\; km\, s^{-1}$).

The lightcurve of He\,II is quite symmetrical with less pronounced
bulges than in C\,III+N\,III, resulting in an eclipse map consisting
of a symmetrical, centred brightness distribution and asymmetric arcs
at different distances from disc centre. The outermost arc (arc 2) is
more easily seen in the eclipse map, while the emission from the 
innermost arc (arc 1) is blendend with that of the central source.
Nevertheless, arc 1 is clearly seen in the asymmetric part of the He\,II
map show in the right panel of Fig.\,\ref{fig2}. The symmetrical emission
component is probably related to the low-velocity component seen in He\,II
Doppler tomograms and is suggested to be due to gas outflow in a wind
emanating  from the inner parts of the disc (see also Marsh and Horne, 1990;
for an alternative interpretation, slingshot prominence from the secondary
star, see Steeghs et~al. 1996). The He\,II arcs contribute about 16 per
cent of the total flux of the eclipse map -- in good agreement with the
results from the Doppler tomography, which indicate that the spirals
contribute $\simeq 15$ per cent of the total He\,II emission (Harlaftis 
et~al. 1999). In comparison, the arcs in the C\,III+N\,III and continuum
maps contribute, respectively, about 30 and 13 per cent of the total flux.

We quantify the properties of the asymmetric arcs by dividing the eclipse
map in azimuthal slices (i.e., `slices of pizza') and computing the
distance at which the intensity is maximum for each azimuth. This
exercice allows to trace the distribution in radius and azimuth of the
spiral structures. The results are plotted in Fig.\,\ref{fig4} as a
function of orbital phase. The diagrams for He\,II were computed from its
asymmetric map and are noisier than those for C\,III+N\,III because most
($\simeq 84$ per cent) of the flux in the eclipse map is subtracted with
the symmetric component. The two spiral shocks are clearly visible in the
intensity diagrams, as well as their distinct locations with respect to 
disc centre. In He\,II the outer spiral (arc2) is brighter than the inner
spiral (arc1), in line with the results of Harlaftis et~al. (1999), while
in C\,III+N\,III arc 1 is brighter than arc 2. The middle panels give
the radial position of the maximum intensity as a function of binary
phase. For C\,III+N\,III, the maximum intensity along arc 1 lies at a
constant distance of $\simeq 0.28\; R_{L1}$ from disc centre while the
maximum intensity of arc 2 occurs at $\simeq 0.55\; R_{L1}$.
The numbers are similar for He\,II.

We computed equivalent Keplerian velocities for each radius assuming
$M_1= 1.0 \pm 0.1 \; M_\odot$ e $R_{\rm L1}= 0.81\; R_\odot$ (Marsh \&
Horne 1990). The results are plotted in the upper panel of Fig.\,\ref{fig4}.
Gray lines show the corresponding uncertainties at the 1-$\sigma$ limit. 
For comparison, the results from the C\,III+N\,III and He\,II Doppler
tomograms (Harlaftis et~al. 1999; see their fig.\,4 for the He\,II
diagram) are shown as dashed lines. Since the C\,III+N\,III Doppler map
is much noisier and blurred than the He\,II Doppler map, the corresponding
diagram is noisier and less reliable than the He\,II diagram on the right
panel. We obtain velocities in the range $850-1050 \; km\, s^{-1}$ for
the spiral 1 compared to the observed $400-770\; km\,s^{-1}$ and in the
range $650-800\; km\,s^{-1}$ for the spiral 2 compared to the observed
$400-550 \; km\,s^{-1}$ (observed values from Harlaftis et~al. 1999). 
The Keplerian velocities calculated from the
radial position of the shocks are systematically larger than those
inferred from the Doppler tomography, suggesting that the gas in the
spiral shocks has sub-Keplerian velocities. This is in line with the
results of the hydrodynamical simulations of Steeghs \& Stehle (1999,
see their fig.\,5), which predicts velocities lower than Keplerian (by as
much as 15 per cent) in the outer disc near the spirals.

We remark that with the white dwarf mass and Roche lobe radius of IP~Peg
the Keplerian velocity at the largest possible disc radius ($R\simeq 0.85\;
R_{\rm L1}$) is about $530 \; km\,s^{-1}$. Therefore, if the observed
velocities of Harlaftis et~al. (1999) do reflect Keplerian motions, 
then the emitting gas should be at the border and even outside the
primary lobe.

Occultation of light from the inner disc regions by the spirals might
produce the out of eclipse variations seen in Fig.\,\ref{fig1}. This is
expected since the spiral waves are also vertically extended. From the
azimuthal position of the spirals in the eclipse map, the maximum
occultation (i.e., the minimum of the orbital modulation) should occur
when the spirals are seen face on, at orbital phases $\simeq -0.3$ and
$\simeq +0.15$ cycle, in agreement with the modulations seen in
Fig.\,\ref{fig1}.
%

\section{Conclusions} \label{fim}

We analyzed eclipse lightcurves of the dwarf novae IP Peg during
the November 1996 outburst in order to confirm the existence, 
constrain the location and investigate the spatial structure of the
spiral shocks observed in the Doppler tomograms.
Our mais results can be summarized as follows:

\begin{itemize}
\item Eclipse maps in the blue continuum and in the C\,III+N\,III
emission line reveal two asymmetric arcs at different azimuth and
radius from disc centre which are consistent with the spiral shocks
seen in the Doppler tomograms. The arcs show an azimuthal extent of
$\sim 90\degr$ and extend from the intermediate to the outer disc 
regions ($R\simeq 0.2 - 0.6\; R_{L1}$). 
The outer radius of the spirals is of the same order of the disc 
outburst radius ($R_{d}\simeq 0.34\;a\simeq 0.6\; R_{L1}$).

\item The He\,II eclipse map is composed of a central brightness source
plus asymmetric arcs at different distances from disc centre. 
The symmetric component probably corresponds to the low-velocity
component seen in He\,II Doppler tomograms and is understood in terms
of gas outflow in a wind emanating from the inner parts of the disc.

\item The spirals contribute about 16 and 30 per cent of the total line
flux, respectively, for the He\,II and C\,III+N\,III lines, and 13 per
cent in the continuum. 

\item The Keplerian velocities derived from the radial position of the
shocks are systematically larger than those inferred from the Doppler
tomography, indicating that the gas in the spiral shocks has
sub-Keplerian velocities.

\end{itemize}

\section*{Acknowledgments}

We thank an anonymous referee for helpful discussions and comments.
This work was partially supported by the PRONEX/Brazil program through the
research grant FAURGS/FINEP 7697.1003.00. RB acknowledges financial 
support from CNPq/Brazil through grant no. 300\,354/96-7.
ETH was supported by the TMR contract ERBFMBICT960971 of the European
Union.

%
\begin{figure*}
 \caption{ Original lightcurves (gray open squares), fitted splines
	(gray dashed lines) and corrected lightcurves (black filled squares),
	for the continuum, C\,III+N\,III and He\,II data. Vertical dotted
	lines mark the ingress/egress phases of the white dwarf for an
	assumed eclipse width of $\Delta\phi=0.086$ cycle (Wood \& Crawford
	1986) and horizontal dotted lines mark the reference (mid-eclipse)
	flux level of the spline fit. }
 \label{fig1}
\end{figure*}

%
\begin{figure*}
 \caption{ Left: Data (dots with error bars) and model (solid lines)
	lightcurves of IP Peg at outburst maximum for the continuum,
	C\,III+N\,III $\lambda 4650$ and He\,II $\lambda 4686$ lines.
	Horizontal dashed lines indicate the uneclipsed component in each
	case. Middle: eclipse maps in a logarithmic grayscale. Right:
	the eclipse maps of the middle panel after subtracting their
	symmetric part; these diagrams emphasize the asymmetric structures.
	Brighter regions are indicated in black; fainter regions in white.
	A cross mark the center of the disc; dotted lines show the Roche 
	lobe and the gas stream trajectory; dotted circles mark disc radii
	of $R=0.2$ and $0.6\;R_{L1}$; the secondary is to the right 
	of each map and the stars rotate counterclockwise. }
\label{fig2}
\end{figure*}

%
\begin{figure*}
\centerline{\psfig{figure=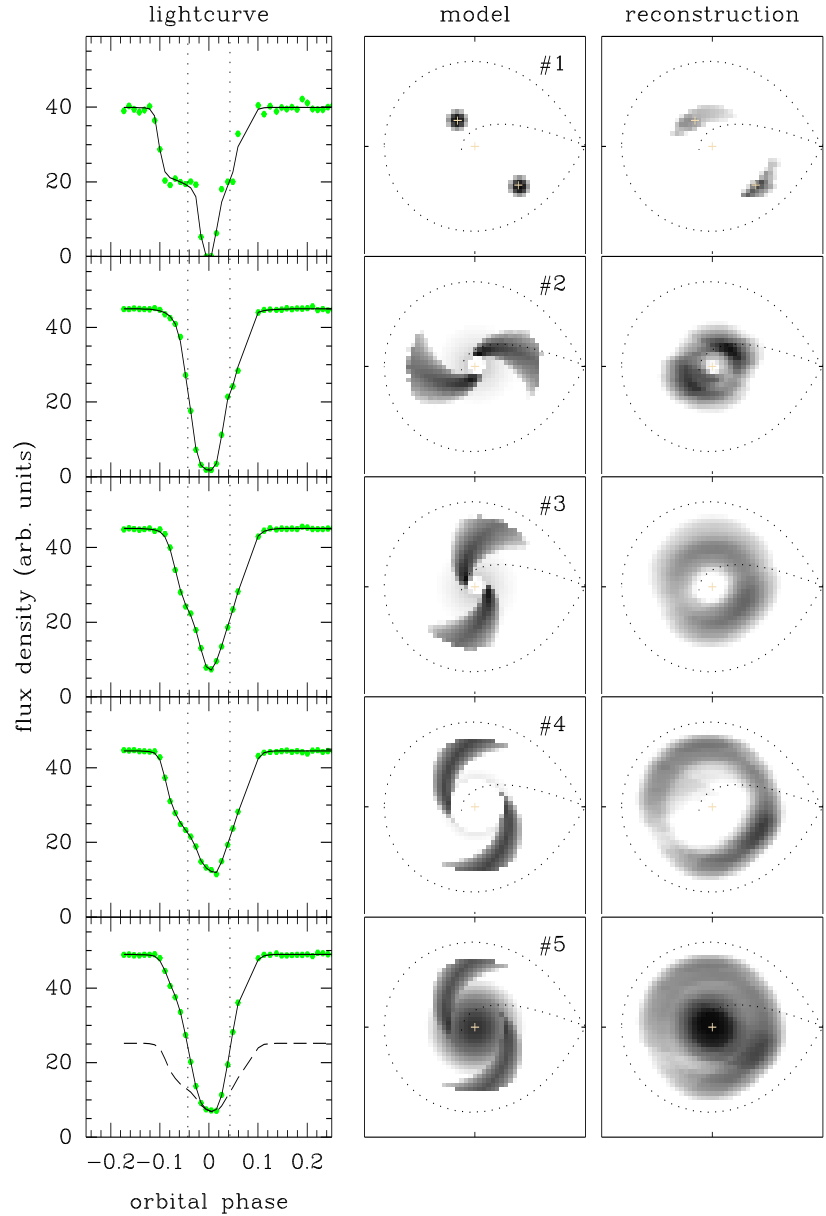,width=14.5cm,rheight=17.5cm}}
\caption{ Simulations with asymmetric brightness distributions.
	Panels in the center show different synthetic images in logarithmic
	grayscale. Panels in the left show noise-added lightcurves derived 
	from these brightness distributions (dots) and the fitted models 
	(solid lines). Vertical dotted lines mark the ingress/egress phases
	of the center of the disc. The dashed curve in the lower panel
	illustrates the contribution from the spiral arms to the eclipse
	shape of model \#5. Panels on the right show the corresponding
	eclipse maps in the same logarithmic grayscale as in the middle panel.
	The notation is the same as in Fig.\,\ref{fig2}. }
\label{fig3}
\end{figure*}

%
\begin{figure*}
 \caption{ The dependency with binary phase of the maximum intensity,
	radius and corresponding Keplerian velocity at maximum intensity,
	as derived from the C\,III+N\,III and He\,II eclipse maps. The
	velocities were computed assuming $M_1= 1.0\pm 0.1\; M_\odot$ and
	$R_{L1}= 0.81\; R_\odot$ (Marsh \& Horne 1990) and the intensities
	are plotted in an arbitrary scale. Gray lines show the uncertainties 
	in the velocity at the 1-$\sigma$ limit. The dependency of velocity
	with orbital phase as derived from the C\,III+N\,III and He\,II Doppler
	tomograms (Harlaftis et al. 1999; see their fig.4) are shown as dashed
	lines for comparison. The location of the spiral arms are indicated by
	horizontal bars with labels 1 (inner spiral) and 2 (outer spiral). }
 \label{fig4}
\end{figure*}

\bsp
\end{document}